\documentclass[preprint,12pt,authoryear]{elsarticle}

\usepackage{amssymb}
\usepackage{amsmath}

\usepackage[utf8]{inputenc}
\usepackage[T1]{fontenc}
\usepackage{lineno}
\usepackage{url}
\usepackage[colorlinks=true,linkcolor=blue,citecolor=blue,urlcolor=blue]{hyperref}
\usepackage{placeins}
\usepackage{xcolor}
\usepackage{soul}
\usepackage{booktabs}

\sethlcolor{yellow}

\journal{Icarus}
\begin{document}

\begin{frontmatter}
		
		\title{Colour changes of Jupiter's Oval BA through microphysical modelling} 
		
		\author[label1,label2]{Asier Anguiano-Arteaga\corref{cor1}} 
		\ead{asier.anguiano@ehu.eus}
		\cortext[cor1]{Corresponding author.}
		
		\author[label1]{Santiago Pérez-Hoyos} 
		\author[label1]{Agustín Sánchez-Lavega} 
		\author[label2]{Patrick G.J. Irwin} 
		
		\affiliation[label1]{organization={Dpto. Física Aplicada, EIB, Universidad del País Vasco UPV/EHU},
			city={Bilbao},
			country={Spain}}
		
		\affiliation[label2]{organization={Department of Physics, Atmospheric, Oceanic and Planetary Physics, University of Oxford},
			city={Oxford},
			country={United Kingdom}}
		
		\begin{abstract}
			Jupiter's Oval~BA undergoes recurrent colour changes whose physical origin remains uncertain. Radiative transfer retrievals indicate that these changes occur in the upper chromophore haze of the vortex annulus, around and above the 0.2--bar level, and are primarily associated with a decrease in optical depth, with no significant change in particle size or haze altitude. We apply a one--dimensional microphysical model to this haze layer, constrained by the retrieved aerosol properties of the red annulus in 2016 and the whiter annulus in 2020, and use it to reproduce the observed colour--change timescale of approximately 0.5 years. Our results indicate that this transition is best reproduced by changes in tropospheric vertical transport within a subsiding annulus, corresponding to preferred downwelling velocities of order 10$^{-4}$--10$^{-3}$~m~s$^{-1}$  at chromophore--bearing pressures. These small vertical velocities may help explain why no clear dynamical signature has yet been identified.
		\end{abstract}
		
		\begin{highlights}
				\item Microphysical and radiative transfer modelling constrains Oval~BA's colour changes
				\item Colour changes in Oval~BA occur on a 0.5-year timescale
				\item The colour--changing annulus is identified as a region of subsidence 
				\item Downwelling velocities at chromophore--bearing pressures are of order 10$^{-4}$--10$^{-3}$ m s$^{-1}$
		\end{highlights}
		
		
\end{frontmatter}
	
\section{Introduction}
\label{Introduction}
\begin{figure}[h]
	\centering
	\includegraphics[width=1\linewidth]{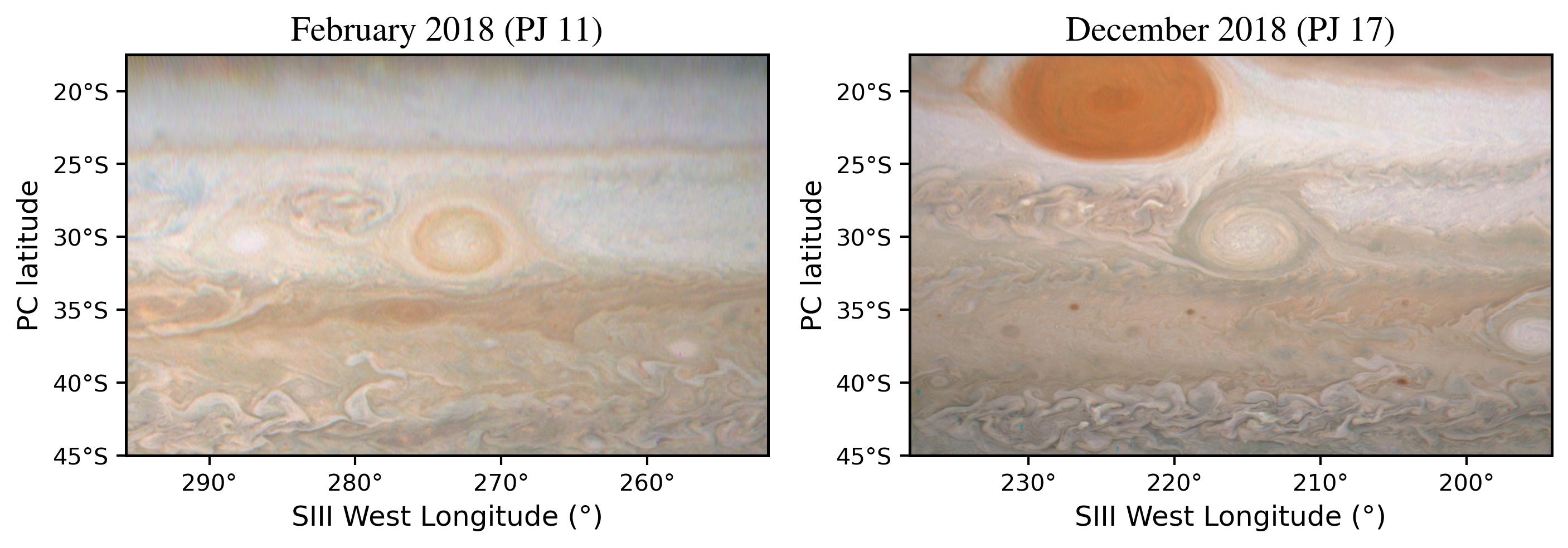}
	\caption{JunoCam views of Oval~BA obtained in February 2018 (left; Perijove~11) and December 2018 (right; Perijove~17), illustrating the colour change of the vortex annulus during 2018. The images were retrieved from the PVOL JunoCam archive \citep{Hueso2018}.}
	\label{fig:OvalBA_colour_change}
\end{figure}

Oval~BA, the second largest and second--longest--lived anticyclone on Jupiter after the Great Red Spot (GRS), formed in 2000 in the South Temperate Belt through the merger of two long--lived white ovals, one of which had itself formed in an earlier merger in 1998 \citep{SanchezLavega1999,SanchezLavega2001}. After remaining white for several years, an annulus surrounding the vortex core turned red in 2005--2006 \citep{SimonMiller2006}. Despite extensive study, later works found no compelling evidence for major changes in either the primary anticyclonic circulation at visible cloud tops or the main vertical cloud structure \citep{Hueso2009,AsayDavis2009,Wong2011}. The reddening timescale was estimated to be about 0.4 years \citep{PerezHoyos2009}. More recently, observations showed that the annulus had reverted to a white state in 2018, as indicated by enhanced short-wavelength reflectivity \citep{SimonWong2024}; this change is illustrated in Fig.~\ref{fig:OvalBA_colour_change}. Using HST/WFC3 imaging and radiative transfer modelling, \citet{AnguianoArteaga2023} showed that this whitening was consistent with a decrease in the optical depth of the upper chromophore layer, with absorption properties compatible with the chromophore proposed by \citet{Carlson2016} and earlier by \citet{FerrisIshikawa1987}. Observations indicate that this whitening was short-lived and was followed by renewed reddening of the Oval~BA annulus, accompanied by a corresponding reduction in short-wavelength reflectivity \citep{SimonWong2024}. This suggests that such colour transitions may recur, although the timing of the observed episodes does not support any clear periodicity. Available HST/WFC3 \citep{MarinelliGreen2025} and JunoCam \citep{Hansen2017} imaging further indicates that the 2018 whitening took place within about 0.6 years, providing a useful observational constraint on the mechanisms responsible for the colour change.

In this work, we apply the microphysical modelling framework recently developed by \citet{AnguianoArteaga2026} for the upper chromophore haze of the GRS to the chromophore layer of Oval~BA. We first identify the combinations of source and transport parameters capable of reproducing the retrieved properties of the vortex before its 2018--2019 whitening. We then introduce controlled perturbations within that parameter space to investigate which changes can reproduce the observed decrease in chromophore optical depth on the relevant timescales, and hence to assess the most plausible mechanisms behind the colour changes of Oval~BA.

\section{Microphysical framework}
\label{sec:microphysics}

We simulate the time--dependent evolution of the upper chromophore haze layer with a one--dimensional microphysical model, based on the computational scheme of \citet{Toon1988} and on our recent implementation for Jupiter’s GRS \citep{AnguianoArteaga2026}.

\subsection{Source, transport and growth}
\label{sec:microphysics_equations}
The evolution of the particle number concentration $C(z,r,t)$ for spherical particles of radius $r$ at altitude $z$ is given by the continuity equation:
\begin{align}
	\frac{\partial C(z,r)}{\partial t} &= C_{\text{inj}}(z,r) \notag \\
	&\quad + \frac{\partial [ K_{zz}(z) \rho] }{\partial z}  \frac{\partial}{\partial z} \left[ \frac{C(z,r)}{\rho} \right]
	- \frac{\partial \left[ W(z,r)  C(z,r) \right]}{\partial z} \notag \\
	&\quad + P_{\text{coag}}(z,r) - L_{\text{coag}}(z,r)
	\label{eq:continuity}
\end{align}

where $\rho$ is the atmospheric gas density, $K_{zz}(z)$ is the eddy--diffusion coefficient, and $W(z,r)$ is the combined transport term for gravitational settling ($w_{\rm fall}$) and background advection:
\begin{equation}
	W(z,r) = w_{\rm fall}(z,r) + w_{\rm trop}
	\label{eq:W}
\end{equation}

In the sign convention used here, positive values of $w_{\rm trop}$ denote upward motion and negative values denote downward motion. The prescribed background velocity $w_{\rm trop}$ is applied only in the troposphere and is set to zero above the tropopause, which we take to be at 0.1 bar, consistent with the strong static stability of the stratosphere associated with the temperature inversion, which suppresses convective vertical motions \citep{Ingersoll2004,CatlingKasting2017}.

The source term $C_{\text{inj}}(z,r)$ in Eq.~\ref{eq:continuity} represents the local net production of newly formed chromophore-bearing particles. It parameterises the conversion of photochemical products into particulate chromophore material, without explicitly modelling the underlying gas-to-particle pathway. Its vertical profile is specified as a Gaussian in pressure, centred at $P_{\rm inj}$ with width set by the FWHM, and with peak injection rate $C_{\rm inj}$:
\begin{equation}
	C_{\rm inj}(P) =
	C_{\rm inj}\,
	\exp\!\left[-\frac{(P-P_{\rm inj})^2}{2\sigma^2}\right],
	\qquad
	\sigma = \frac{\mathrm{FWHM}}{2\sqrt{2\ln 2}}
	\label{eq:inj}
\end{equation}

In Eq.~(\ref{eq:continuity}), $P_{\rm{coag}}$ and $L_{\rm{coag}}$ represent the production of particles of size $r$ at a given altitude by coagulation of smaller particles and their loss due to coagulation with other particles, respectively. Changes in the concentration of particles at a given altitude and size bin due to vertical transport are accounted for separately by the eddy--diffusion and $W(z,r)$ terms. Coagulation is computed with a kernel $K_{\rm coag}$, for which we adopt the formulation given by \citet{SitarskiSeinfeld1977}. Particle growth is limited by electrostatic repulsion between like--charged particles, parameterised through a sticking efficiency $\alpha_s$ \citep{Pollack1987},
\begin{equation}
	K_{\rm coag} \rightarrow \alpha_s\,K_{\rm coag}, \quad
	\alpha_s = \exp\!\left[-\frac{k_C\,q_1 q_2}{k_B T(r_1+r_2)}\right], \quad
	q_i = (Q\,r_i)\,e
	\label{eq:alpha}
\end{equation}
where $k_C$ is the Coulomb constant, $k_B$ is Boltzmann’s constant and $T$ is the temperature. Particle charge $q_i$ is expressed through a charge--per--radius factor $Q$ (electrons~$\mu$m$^{-1}$), with $r_i$ in $\mu$m and $e$ the electron charge. Other material-dependent effects, such as surface energy and particle morphology, may also influence the cohesion and growth of real haze particles \citep{Yu2020}, but are not explicitly included here since they are not constrained for Jovian chromophore materials.

The atmospheric inputs required by Eqs.~(\ref{eq:continuity})--(\ref{eq:alpha})---namely the background P--T structure, the eddy--diffusion profile $K_{zz}(z)$ and the settling velocity $w_{\rm fall}(z,r)$---are detailed in Supplementary Text~S1 and Fig.~S1.


\subsection{Target aerosol properties}
\label{RT_constraints}

Reference constraints for Oval~BA's upper chromophore layer are taken from the HST/WFC3 radiative transfer retrievals reported by \citet{AnguianoArteaga2023}. Those retrievals provide aerosol properties for a reddish annulus in 2016 and a whitish annulus in 2020. The 2016 constraint corresponds to a single HST/WFC3 observing run obtained on 11--12 December 2016 under program \href{https://www.stsci.edu/hst-program-info/program/?program=14661}{GO~14661}. The 2020 constraint is based on the average of retrievals from three HST/WFC3 observing runs, obtained on 22--23 July 2020 under program \href{https://www.stsci.edu/hst-program-info/program/?program=16053}{GO~16053}, and on 25 August and 20 September 2020 under program \href{https://www.stsci.edu/hst-program-info/program/?program=15929}{GO/DD~15929}. The retrieved $r_{\rm eff}$ and $\tau(900~{\rm nm})$ values were obtained from multi--filter HST/WFC3 spectra spanning the UV to the near--IR; the corresponding filter sets for each observing run are reported in \citet{AnguianoArteaga2023}.

In particular, simulations are first required to reproduce the 2016 effective radius and optical depth of the Oval~BA annulus, with target ranges $r_{\rm eff}=0.2$--0.3~$\mu$m and $\tau(900~{\rm nm})=1.0$--1.4, computed above 0.2~bar from the microphysical output. The transition to the 2020 state is represented as a 30--35\% reduction in $\tau(900~{\rm nm})$. Reported changes in $r_{\rm eff}$ between 2016 and 2020 were not significant within the retrieval uncertainties; accordingly, $r_{\rm eff}$ is required to remain within the original target range. In addition, the simulations are evaluated against the inferred observational limit that the whitening took place within $\sim$0.6~years. Optical depth is computed using the complex refractive indices retrieved by \citet{AnguianoArteaga2023}.
		
\subsection{Numerical configuration and free parameters}
\label{sec:num_config}

The model domain spans $P = 1$--$10^{-3}$~bar (0--150 km) and is discretised into 51 layers. The particle size distribution is represented using 28 geometrically spaced radius bins spanning 0.01--5.12~$\mu$m. Convergence to steady state is identified by requiring $<0.5$\% variability in both effective radius and optical depth over five consecutive outputs, sampled every 1,000 time steps. Simulations are first run for $\sim$15~Earth years; cases matching the target ranges for the 2016 state are then re--run for twice the duration, with a parameter change applied at the midpoint (primarily in $C_{\rm inj}$ and $w_{\rm trop}$), and allowed to relax to a new steady state.

Baseline simulations aimed at reproducing Oval~BA in its 2016 reddish--annulus state explore the free parameters summarised in Table~\ref{tab:param_space}. Two dynamical scenarios are considered: Scenario~A, with tropospheric upwelling ($w_{\rm trop}>0$), and Scenario~B, with tropospheric downwelling ($w_{\rm trop}<0$).  For Scenario~A (upwelling), the lower bound of $10^{-5}$~m~s$^{-1}$ for $w_{\rm trop}$ is obtained by scaling the \citet{Conrath1981} estimate for the GRS to Oval~BA using the smaller temperature departure from its surroundings ($\Delta T \sim 2$~K; \citealt{Cheng2008}). Values close to the $w_{\rm trop}$ upper bound no longer reach steady state within the required time. For Scenario~B (downwelling), we adopt the same minimum magnitude as in Scenario~A, and take four times the estimate by \citet{dePater2010} as an upper bound to explore strong subsidence. For each dynamical scenario, the $C_{\rm inj}$ bounds were chosen so that the valid solutions lie within the sampled bracket. The discrete values of $P_{\rm inj}$ are motivated by the chromophore levels (P $\leq$ 0.1 bar) inferred in \citet{AnguianoArteaga2023} and by the expected formation region of the \citet{Carlson2016} chromophore (i.e., where C$_2$H$_2$ and NH$_3$ co--exist), with corresponding FWHM fractions that span from narrow to broad injection profiles. Valid solutions are obtained for $Q$ values above the lower bound explored, and the upper bound is guided by the larger charging inferred by \citet{Moreno1996} for Jupiter’s south polar aerosols.

\begin{table}[t]
	\centering
	\caption{Summary of the explored parameter space. For $C_{\rm inj}$ and $w_{\rm trop}$, brackets indicate the lower and upper bounds of the explored ranges. Scenario~A corresponds to upwelling ($w_{\rm trop}>0$) and Scenario~B to downwelling ($w_{\rm trop}<0$). A total of 3,840 simulations were performed (1,920 per scenario).}
	\label{tab:param_space}
	\resizebox{\linewidth}{!}{%
	\begin{tabular}{lcccc}
		\toprule
		Parameter & Symbol &
		\begin{tabular}[c]{@{}c@{}}Explored\\values\end{tabular} &
		Sampling &
		\begin{tabular}[c]{@{}c@{}}Introduced\\in\end{tabular} \\
		\midrule
		Peak injection rate (Sc.~A) & $C_{\rm inj}$ &
		$[10^{-3},\,1]$ particles~cm$^{-3}$~s$^{-1}$ &
		$N=8$, log spaced & Eq.~(\ref{eq:inj}) \\
		Peak injection rate (Sc.~B) & $C_{\rm inj}$ &
		$[10^{-2},\,10]$ particles~cm$^{-3}$~s$^{-1}$ &
		$N=8$, log spaced & Eq.~(\ref{eq:inj}) \\
		Tropospheric velocity (Sc.~A) & $w_{\rm trop}$ &
		$[10^{-5},\,10^{-3}]$ m~s$^{-1}$ &
		$N=8$, log spaced & Eq.~(\ref{eq:W}) \\
		Tropospheric velocity (Sc.~B) & $w_{\rm trop}$ &
		$[-10^{-5},\,-8.0\!\times\!10^{-2}]$ m~s$^{-1}$ &
		$N=8$, log spaced & Eq.~(\ref{eq:W}) \\
		Peak injection pressure & $P_{\rm inj}$ &
		$\{0.1,\,0.2\}$ bar &
		$N=2$ & Eq.~(\ref{eq:inj}) \\
		Injection width & FWHM &
		$\{0.3,\,0.6,\,0.9\}\,P_{\rm inj}$ &
		$N=3$ & Eq.~(\ref{eq:inj}) \\
		Charge per radius & $Q$ &
		$\{10,\,15,\,20,\,25,\,30\}$ electrons~$\mu$m$^{-1}$ &
		$N=5$ & Eq.~(\ref{eq:alpha}) \\
		\bottomrule
	\end{tabular}
	}
\end{table}

\section{Results}
\label{results}	
A summary of Scenario~A simulations that steadily reach the target ranges for Oval~BA in 2016 is provided in Supplementary Table~S1. Valid upwelling velocities span $1.0\!\times\!10^{-5}$ to $1.4\!\times\!10^{-4}$~m~s$^{-1}$, with convergence times exceeding 7~Earth years. The associated column--integrated mass injection fluxes span $\sim$1$\times\!10^{-12}$ to 7$\times\!10^{-12}$~kg~m$^{-2}$~s$^{-1}$.

\begin{figure}[h]
	\centering
	\includegraphics[width=\linewidth]{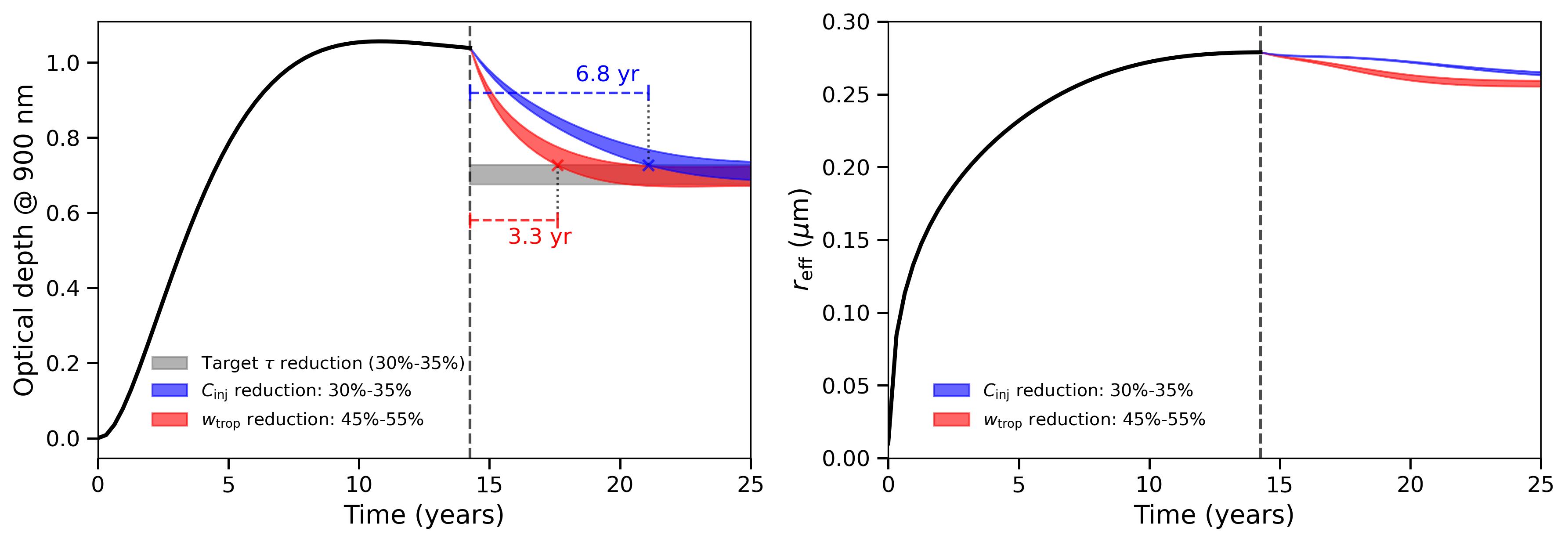}
	\caption{Response of $\tau(900~\mathrm{nm})$ (left) and $r_{\rm eff}$ (right) to imposed parameter changes at $t\approx15$~years (dashed line) for a Scenario~A (upwelling) baseline case with $w_{\rm trop}=1.4 \times 10^{-4}$ m s$^{-1}$ (marked case in Table S1). Blue: $C_{\rm inj}$ reduced by 30--35\%. Red: $w_{\rm trop}$ reduced by 45--55\%. The grey band marks the target 30--35\% decrease in optical depth between 2016 and 2020, and the coloured guides indicate the elapsed time from the imposed perturbation to first reaching this band.  In all cases, $r_{\rm eff}$ exhibits only minor variations and remains within the target $0.2$--$0.3~\mu$m range.}
	\label{fig:scenarioA_changes}
\end{figure}

Fig.~\ref{fig:scenarioA_changes} illustrates the response of a Scenario~A steady state to parameter changes applied at mid--integration ($t\approx15$~years). In particular, it corresponds to the valid simulation that reaches the target optical--depth decrease in the shortest time when $w_{\rm trop}$ is varied. Reducing either $C_{\rm inj}$ or $w_{\rm trop}$ reproduces the required 30--35\% decrease in $\tau(900~\mathrm{nm})$, while keeping $r_{\rm eff}$ within the target range. However, the adjustment is significantly faster when reducing $w_{\rm trop}$, with the target $\tau(900~\mathrm{nm})$ range reached after $\sim$3~years, compared with $\sim$7~years for the $C_{\rm inj}$ reduction. Optical depth can also be decreased by reducing $Q$, which enhances coagulation and increases the drainage of mass by sedimentation ($w_{\rm fall}>w_{\rm trop}$ for larger particles), but this increases $r_{\rm eff}$ by $\sim$ 40\% and drives it outside the target range. Such particle--size changes have not been reported during previous reddening and whitening episodes \citep{PerezHoyos2009,AnguianoArteaga2023}. Exploring the different injection configurations corresponding to valid simulations with $w_{\rm trop}=1.4\times10^{-4}$~m~s$^{-1}$, the required reduction in $w_{\rm trop}$ ranges from $\sim$40\% to 80\%. Conversely, baseline cases with $w_{\rm trop}<1.4\times10^{-4}$~m~s$^{-1}$ do not reach a $\geq$30\% decrease in optical depth even for a 100\% reduction in $w_{\rm trop}$, i.e., the limiting case of zero upwelling. Across all valid Scenario~A solutions (Table~S1), the shortest time to reach the target decrease in optical depth is 3.3~years when varying $w_{\rm trop}$ (Fig.~\ref{fig:scenarioA_changes}), while the corresponding minimum time for variations in $C_{\rm inj}$ is 3.5~years and is obtained for a different simulation with $w_{\rm trop}=10^{-5}$~m~s$^{-1}$. Although Scenario~A can therefore reproduce the required optical--depth decrease, its response remains too slow compared with the observed sub--annual colour--change timescale.

Scenario~B simulations that steadily reproduce the 2016 Oval~BA target ranges are listed in Supplementary Table~S2. Valid downwelling velocities span $-3.6\!\times\!10^{-5}$ to $-8\!\times\!10^{-2}$~m~s$^{-1}$, reaching the strongest subsidence explored in our parameter space. Scenario~B yields shorter convergence times than Scenario~A, with values down to approximately 4~Earth years. The associated column--integrated mass injection fluxes span $\sim 2\!\times\!10^{-12}$ to $5\!\times\!10^{-11}$~kg~m$^{-2}$~s$^{-1}$. These values extend to larger injections than in the upwelling case, since downwelling more efficiently clears the upper troposphere and therefore requires stronger replenishment to maintain the haze layer. However, much of the flux range overlaps with that obtained for Scenario~A.

\begin{figure}[h]
	\centering
	\includegraphics[width=\linewidth]{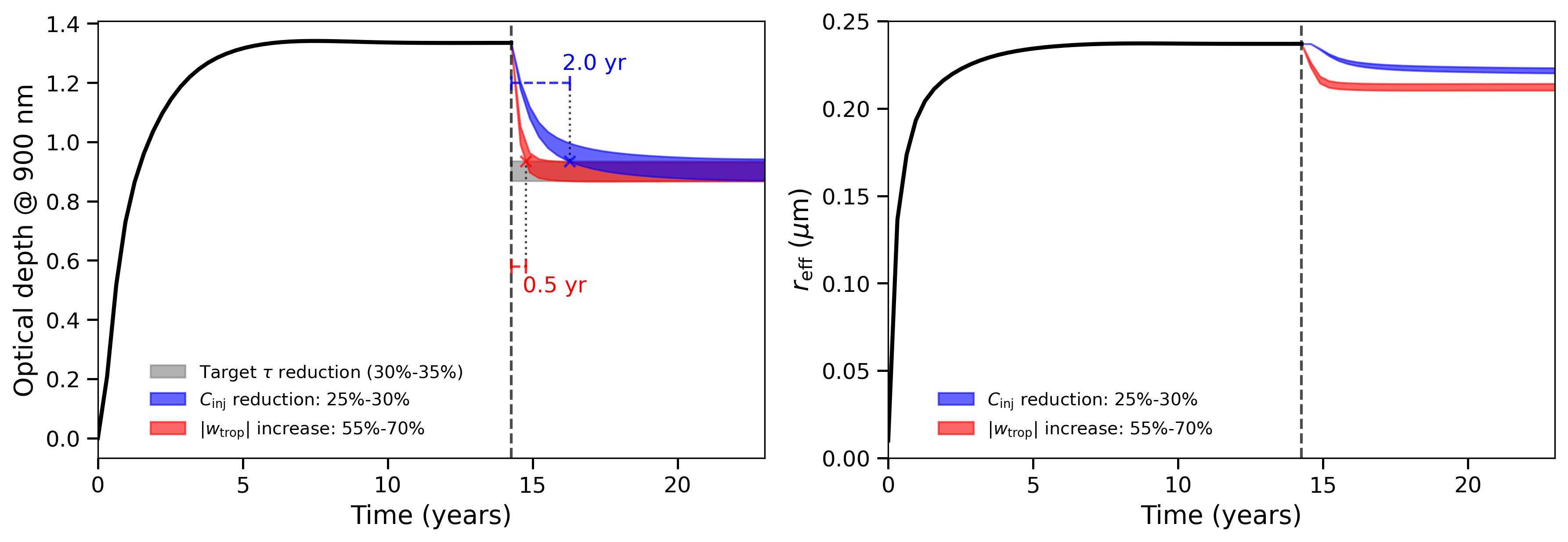}
	\caption{Response of $\tau(900~\mathrm{nm})$ (left) and $r_{\rm eff}$ (right) to imposed parameter changes at $t\approx15$~years (dashed line) for a Scenario~B (downwelling) baseline case with $w_{\rm trop}=-4.7\times10^{-4}$~m~s$^{-1}$ (marked case in Table~S2). Blue: $C_{\rm inj}$ reduced by 25--30\%. Red: $|w_{\rm trop}|$ increased by 55--70\%. The grey band marks the target 30--35\% decrease in optical depth between 2016 and 2020, and the coloured guides indicate the elapsed time from the imposed perturbation to first reaching this band. In all cases, $r_{\rm eff}$ exhibits only minor variations and remains within the target $0.2$--$0.3~\mu$m range.}
	\label{fig:scenarioB_changes}
\end{figure}

A first Scenario~B sensitivity sweep was performed by strongly increasing the downwelling magnitude, $|w_{\rm trop}|$, by a factor of 10 in all 39 valid steady--state solutions. The outcome is controlled mainly by the baseline $|w_{\rm trop}|$: only cases with $|w_{\rm trop}| \leq 4.7\times10^{-4}$~m~s$^{-1}$ reach or exceed the required decrease in $\tau(900~\mathrm{nm})$, whereas none with $|w_{\rm trop}| \geq 1.7\times10^{-3}$~m~s$^{-1}$ do so. Fig.~\ref{fig:scenarioB_changes} illustrates the response of a Scenario~B steady state with baseline $w_{\rm trop}=-4.7\!\times\!10^{-4}$~m~s$^{-1}$ to parameter changes applied at mid--integration ($t\approx15$~years). As for Scenario~A, we again show the valid simulation that reaches the target optical--depth decrease in the shortest time when $w_{\rm trop}$ is varied. Reducing $C_{\rm inj}$ or increasing the downwelling magnitude, $|w_{\rm trop}|$, both reproduce the required 30--35\% decrease in $\tau(900~{\rm nm})$ while keeping $r_{\rm eff}$ within the target range. By contrast, changes in $Q$ again produce much larger shifts in $r_{\rm eff}$, moving it outside the observed range. As in the upwelling case, the adjustment is significantly faster when increasing $|w_{\rm trop}|$ than when changing $C_{\rm inj}$, with the target $\tau(900~{\rm nm})$ range reached after $\sim$0.5 years for the $w_{\rm trop}$ perturbation, compared with $\sim$2.0 years for the $C_{\rm inj}$ reduction. Across all valid Scenario~B solutions (Table~S2), the shortest time to reach the target decrease in optical depth is 0.5~years when varying $w_{\rm trop}$ (Fig.~\ref{fig:scenarioB_changes}), while the corresponding minimum time for variations in $C_{\rm inj}$ is 1.2~years and is obtained for a different simulation with $w_{\rm trop}=-1.7\times10^{-3}$~m~s$^{-1}$. 

\section{Discussion}
\label{discussion}
The typical timescale of the Oval BA colour changes is not perfectly constrained. Although the vortex can be monitored quasi--continuously through amateur observations, differences in image processing applied to those datasets make quantitative colour comparisons difficult. For the first observed colour transition of Oval~BA, \citet{PerezHoyos2009} estimated a reddening timescale of about 0.4 years. The available HST/WFC3 and JunoCam imaging indicates that the 2018 whitening also took place on a sub--annual timescale. In particular, the transition from an annulus with an intermediate reddish colour to a whitish annulus appears to have occurred over about 0.25 years, from October to December 2018, while the full evolution from a clearly reddish annulus to that same whitish state was completed within about 0.6 years, from April to December 2018. These observational limits are broadly consistent with the estimate of \citet{PerezHoyos2009}, although an appreciable visual change need not imply that the atmosphere has already reached its new steady state. Our results indicate that the timescale of the Oval BA annulus colour change is better reproduced by local downwelling than by upwelling. In the upwelling scenario, the target optical--depth decrease is not reached in less than 3 years in any case. Likewise, only changes in downward vertical velocity bring the response close to the observationally inferred sub--annual timescale, with our shortest timescale being 0.5 years. Opposite parameter changes, corresponding to reddening rather than whitening, occur on very similar timescales (Fig.~S2). This suggests that dynamical variations in vertical transport better reproduce the characteristic timescales of Oval BA colour changes.

Our results are broadly consistent with the secondary--circulation picture of Oval~BA proposed by \citet{dePater2010}, \citet{Wong2011} and \citet{Marcus2013}. In that framework, air rises over the vortex core and returns downward near the red annulus. Using a nearly divergence--free scaling, $v_r \approx wL/D$, with $L=$3000~km and $D/L=0.02$, \citet{dePater2010} inferred a characteristic radial velocity of $\sim$1~m~s$^{-1}$ and a characteristic vertical velocity scale of $\sim2\times10^{-2}$~m~s$^{-1}$. This estimate should not be interpreted as a direct measurement of the local vertical velocity in the chromophore--bearing annulus. Our model instead constrains an effective vertical transport velocity, $w_{\rm trop}$, most directly relevant to the pressure levels where the chromophore layer is located, around and above the 0.2--bar level. In Scenario~B, the observed decrease in annulus optical depth is reproduced by increasing the magnitude of the local downwelling from values up to $|w_{\rm trop}|\sim5\times10^{-4}$~m~s$^{-1}$ in the red state to approximately $10^{-3}$~m~s$^{-1}$ as the vortex evolves towards a white state. These values refer to the local downwelling velocity before and after the imposed change, not to the perturbation itself. They are well below the characteristic velocity scale of \citet{dePater2010}, consistent with an effective transport velocity near the chromophore layer and close to the stably stratified tropopause. Since $w_{\rm trop}$ is not equivalent to the characteristic vertical velocity scale of the full secondary circulation, we do not use it to infer a corresponding radial velocity. We instead note that, at the pressure levels relevant to the chromophore layer, the colour transition requires a change between already weak effective downwelling velocities. This does not imply that the flow is confined to the chromophore layer; rather, our radiative--transfer and microphysical constraints are most sensitive to chromophore--bearing pressures, so any associated changes at cloud level or in the thermal field, if present, need not be large. Such changes would therefore be difficult to detect in existing imaging or cloud--tracking data. This may help explain why the reddening of Oval~BA was not accompanied by clearly measurable changes in either its overall velocity field or its upper cloud vertical structure \citep{AsayDavis2009,Hueso2009,PerezHoyos2009}.

Under this picture, the inner region of Oval~BA would correspond to the upwelling branch of the secondary circulation, whereas the red annulus would trace its subsiding branch. Thus, we also considered the central region using the upwelling parameter grid from Scenario~A, but now adopting a target range of $\tau(900\,\mathrm{nm})=$ 0.7--0.8 based on the values reported by \citet{AnguianoArteaga2023} for the central region of the oval. In this context, it is worth recalling that the difference between a red and a white state does not imply the absence of chromophore-bearing aerosol in the latter, but rather a lower optical depth, and therefore a weaker contribution of the chromophore to the observed short-wavelength absorption. The solutions obtained for the central region yield column--integrated mass injection fluxes of $\sim$7$\times$10$^{-13}$ to 2.4$\times$10$^{-12}$~kg~m$^{-2}$~s$^{-1}$ and positive, upwelling velocities between 10$^{-5}$ and 1.4$\times$10$^{-4}$~m~s$^{-1}$. These mass injection values are generally lower than those inferred for the subsiding red annulus. This contrast is consistent with a scenario in which material supplied over the upwelling core is redistributed towards the annulus by the secondary circulation, where it subsequently undergoes local subsidence and processing.

In addition, as in the GRS microphysical modelling of \citet{AnguianoArteaga2026}, the material injection rates required by the preferred Oval~BA solutions appear too large to be explained solely by the local production budget of the \cite{Carlson2016} chromophore from C$_2$H$_2$ and photodissociated NH$_3$. This suggests that the aerosol responsible for the observed spectra may contain the chromophore proposed by \citet{Carlson2016}, whose optical properties reproduce the observed spectra remarkably well \citep{Baines2019,Braude2020,AnguianoArteaga2023}, mixed with other non--colouring material.

An alternative interpretation builds on the global climate--change scenario proposed by \citet{Marcus2004}, in which the mergers of the white ovals that ultimately formed Oval~BA were predicted to produce a broader temperature change on Jupiter. Applied to Oval~BA, this scenario suggested that a modest hemispheric or global temperature increase could have helped expose previously hidden red chromophores in the warm annulus \citep{dePater2010,Wong2011,Marcus2013}. However, both for Oval~BA \citep{AnguianoArteaga2023} and the GRS, the chromophore--bearing haze is inferred to reside at high altitude, around or above the 0.2--bar level, i.e. above the main ammonia cloud deck \citep{Baines2019,Braude2020,AnguianoArteaga2021}. At these levels, the visibility of the chromophore would not be directly coupled to the phase state of ammonia in the deeper cloud. Furthermore, the zonal--mean upper--tropospheric temperature contours shown by \citet{Fletcher2016} do not reveal any obvious large difference at the latitude of Oval~BA between the 2000 and 2014 epochs, despite the different colour states of the vortex.

Our ability to discriminate between upwelling and downwelling relies on the timescale of the pronounced colour changes of Oval~BA. By contrast, this analysis cannot be applied to the GRS in the same way, since it does not exhibit similarly strong colour transitions on comparably well--defined timescales, and its secondary circulation therefore cannot be evaluated through our modelling in the same manner. Nevertheless, thermal and compositional studies of the GRS do suggest subsidence outside the red core, as indicated by the low aerosol opacity and enhanced 5--$\mu$m emission in the area surrounding the vortex \citep{Fletcher2010,Harkett2024}. More generally, the present modelling is applicable to elevated anticyclones whose reddish colouring is associated with a chromophore--bearing haze near or above the $\sim$0.2--bar level. It is therefore not directly transferable to Jupiter's belts and zones. Although belts are also commonly linked to subsidence, they are deeper atmospheric regions, their colouration is more brownish than reddish, and their appearance is associated with a deeper cloud and aerosol structure than the detached upper haze layer considered here \citep{Fletcher2020}. In that sense, our results are most relevant to high--altitude red anticyclones, rather than to the general belt--zone circulation of Jupiter.

\section{Conclusions}
\label{conclusions}
Our results support a picture in which the annulus of Oval~BA, where the main colour changes occur, is a region of downwelling rather than upwelling. The preferred solutions reproduce the observed whitening without requiring significant changes in either haze altitude or effective radius. Upwelling solutions do not reproduce the required decrease in optical depth on timescales shorter than about 3 years. Within a subsiding annulus, changes in vertical transport provide a much better match to the observationally inferred sub--annual timescale of the colour change, broadly consistent with the earlier $\sim$0.4-year estimate of \citet{PerezHoyos2009}. The weak vertical motions implied by the preferred solutions, which are most relevant to chromophore-bearing pressure levels, and their limited variations, are also consistent with the absence of detectable dynamical changes at cloud level. Further assessment of this interpretation would therefore require dedicated dynamical modelling capable of reproducing the vortex secondary circulation in a self-consistent way.

\section*{Acknowledgments}		
This work was supported by the Basque Government (Grupos de Investigación, IT1742-22), Elkartek KK-2025/00106 and by Grant PID2023--149055NB--C31 funded by MICIU/AEI/10.13039/501100011033 and by FEDER, UE.

A.~Anguiano-Arteaga was supported by the \textit{Programa de Perfeccionamiento de Personal Investigador Doctor 2024--2027} of the Basque Government.

\end{document}